\documentclass[fleqn,usenatbib]{mnras}

\usepackage{newtxtext,newtxmath}

\usepackage[T1]{fontenc}

\DeclareRobustCommand{\VAN}[3]{#2}
\let\VANthebibliography\thebibliography
\def\thebibliography{\DeclareRobustCommand{\VAN}[3]{##3}\VANthebibliography}

\usepackage{tabularx}
\usepackage{multirow}
\usepackage{hyperref}
\usepackage{makecell}
\usepackage{xcolor}
\usepackage{physics}
\usepackage[normalem]{ulem}
\usepackage{bbold}

\usepackage{graphicx}	
\usepackage{amsmath}	

\newcolumntype{Y}{>{\centering\arraybackslash}X}
\newcolumntype{W}[1]{>{\centering\arraybackslash}p{#1}}

\defcitealias{fiorilli_hmf_2026}{F26}

\title[{\sc Aletheia:} Emulating the HMF with evolution mapping]{\textsc{Aletheia}: Emulating the halo mass function with evolution mapping}

\author[A. Fiorilli et al.]{
Andrea Fiorilli,$^{1}$\thanks{E-mail: fiorilla@mpe.mpg.de (AF)}
Ariel G. Sánchez,$^{1,2}$
Andrés N. Ruiz,$^{3,4}$
and Facundo Rodriguez$^{3,4}$
\\
$^{1}$Max-Planck-Institut f\"ur extraterrestrische Physik, Postfach 1312, Giessenbachstrasse 1, 85748 Garching, Germany\\
$^{2}$Universit\"ats-Sternwarte M\"unchen,  Fakult\"at f\"ur Physik, Ludwig-Maximilians-Universit\"at M\"unchen, Scheinerstrasse 1, 81679 M\"unchen, Germany\\
$^{3}$Instituto de Astronomía Teórica y Experimental, CONICET-UNC, Laprida 854, X5000BGR, Córdoba, Argentina\\
$^{4}$Observatorio Astronómico, Universidad Nacional de Córdoba, Laprida 854, X5000BGR, Córdoba, Argentina\\
}

\date{Accepted XXX. Received YYY; in original form ZZZ}

\pubyear{\the\year{}}

\begin{document}
\label{firstpage}
\pagerange{\pageref{firstpage}--\pageref{lastpage}}
\maketitle

\begin{abstract}
We present an emulator of the halo mass function (HMF) based on Gaussian-process regression, that exploits the evolution mapping framework.
This framework splits cosmological parameters into shape parameters ($\vb{\Theta}_\mathrm{s}$), which set the shape of the linear power spectrum, and evolution parameters ($\vb{\Theta}_\mathrm{e}$), which only rescale its amplitude, with their net effect entirely captured by the clustering amplitude, $\sigma_{12}$. 
The emulation acts in two steps: first, an emulator predicts the overall shape of the HMF given $\vb{\Theta}_\mathrm{s}$ and $\sigma_{12}$; then, a second emulator applies a small correction accounting for the residual dependence on the integrated growth history of density fluctuations.
The emulator is trained on a suite of 100 simulations of flat $\Lambda$CDM cosmologies but, owing to evolution mapping, can be applied to predict the abundance of haloes in cosmologies with any extended space of $\vb{\Theta}_\mathrm{e}$, such as curvature, quintessence and dynamical dark energy.
We test the emulator against independent $\Lambda$CDM runs and on an additional validation suite of dynamical dark energy simulations, achieving per cent-level accuracy over a wide range of masses and redshifts, with the performance only slightly degrading at early times, where the training data points are fewer.
Compared to other state-of-the-art HMF emulators, its accuracy is higher in nearly all tested cases.
Our results demonstrate the broad applicability of the evolution mapping framework to build very accurate emulators of statistics of the density field.
The emulator is publicly available as part of the \textsc{Aletheia} emulator suite.
\end{abstract}




\section{Introduction}
\label{sec:intro}

The abundance of dark matter haloes as a function of their mass, described by the halo mass function (HMF), is a crucial ingredient of the theory of large-scale structure (LSS).
In the standard cosmological picture, galaxies form and evolve within haloes \citep{White_Rees_1978}, so that the statistics of the observed galaxy distribution are tightly connected to those of the underlying halo population, as described, e.g., by halo occupation distribution models \citep{berlind_weinberg_hod_2002, zheng_hod_2005} or by the halo model of LSS \citep[][see \citealt{asgari_halo_model_2023} for a recent review]{cooray_sheth_halo_model_2002}.
The HMF is also the central theoretical quantity of cluster-abundance cosmology: the number counts of galaxy clusters as a function of mass and redshift are a sensitive probe of the growth of structure and of the dark energy equation of state \citep[e.g.,][]{haiman_cluster_counts_2001, Evrad_2002, allen_cluster_cosmology_review_2011}.
Ongoing and upcoming surveys such as \textit{eROSITA} \citep{predehl_erosita_instrument_2021}, \textit{Euclid} \citep{laureijs_euclid_red_book_2011, euclid_overview_2024}, DESI \citep{DESI_2016} and LSST \citep{LSST_2019} are delivering cluster and galaxy samples of unprecedented statistical power \citep[see, e.g., the forecasts of][]{sartoris_euclid_clusters_2016}, and the first of these samples already provide competitive cosmological constraints \citep[e.g.,][]{bocquet_spt_des_clusters_2024, ghirardini_erass1_cosmology_2024, DES_Y3_clusters_2025}.
Fully exploiting these data sets requires accurate theoretical predictions of the halo abundance across wide ranges of mass, redshift and cosmological parameters.

The first analytical description of the HMF dates back to \citet{press_schechter_hmf_1974}, who combined the spherical collapse model \citep{gunn_gott_spherical_collapse_1972} with the Gaussian statistics of the linear density field.
This approach was later given firmer theoretical bases by the excursion set formalism \citep{Peacock_Heavens_1990, bond_excursion_set_1991, Lacey_Cole_1993} and extended to ellipsoidal collapse \citep{Sheth_Tormen_1999, sheth_mo_tormen_hmf_2001}; see \citet{zentner_excursion_set_review_2007} for a review.
A common feature of these models is that the HMF is \emph{universal}: all the dependence on cosmology, redshift and halo mass is encoded in a single variable, the peak height $\nu$, through a multiplicity function $f(\nu)$ of fixed functional form.
However, with N-body simulations reaching higher resolutions, it has become clear that universality only holds approximately: the multiplicity function exhibits residual dependencies on redshift and cosmology at the level of ten per cent or more \citep[e.g.,][]{jenkins_hmf_2001, reed_hmf_2007, tinker_hmf_2008, crocce_hmf_2010, bhattacharya_hmf_2011, Despali_2016, diemer_hmf_2020}.
Several fitting functions calibrated on simulations model the deviations from universality empirically, typically through redshift-dependent parameters \citep[e.g.,][]{warren_hmf_2006, tinker_hmf_2008, courtin_hmf_2011, watson_hmf_2013, bocquet_hmf_2016}, but their accuracy degrades at high redshift or in non-standard cosmologies.
More recent studies traced the physical origin of the non-universality to two ingredients: the shape of the linear matter power spectrum, $P_\mathrm{L}(k)$, at the scales relevant for halo formation, and the history of structure formation \citep[][hereafter F26]{ondaromallea_hmf_2021, castro_euclid_hmf_2023, fiorilli_hmf_2026}.
In particular, \citetalias{fiorilli_hmf_2026} provided a physically motivated description of the growth-history dependence within the evolution mapping framework \citep{sanchez_arguments_against_h_2020, sanchez_evo_mapping_2022}, resulting in a fitting function that remains accurate in cosmologies with curvature, quintessence and dynamical dark energy.

In the last few years, an alternative approach has emerged.
Instead of relying on analytical prescriptions, one can exploit machine learning to build an \textit{emulator} that learns the HMF directly from N-body simulations.
Emulators are statistical models trained to reproduce the output of expensive numerical simulations across a wide range of input parameters: rather than running a full simulation for each new set of cosmological parameters, an emulator interpolates across its training set and returns predictions at negligible computational cost.
Provided that the training simulations have sufficiently high resolution and cover a sufficiently wide region of parameter space, this strategy yields accurate predictions of the HMF without the need to assume a specific functional form, as demonstrated by the numerous HMF emulators released in recent years \citep{nishimichi_darkquest_emulator_2019, mcclintock_aemulus_hmf_emulator_2019, bocquet_miratitan_hmf_emulator_2020, saez_emantis_hmf_emulator_2024, shen_aemulusnu_hmf_emulator_2025, chen_yu_CSST_emulator_2025, buisman_nn_hmf_2025}.
Gaussian process regression and neural networks are the most commonly used frameworks for this purpose: the former natively provides an estimate of the emulation uncertainty through its posterior variance, while the latter scale better to large training sets.
The main limitation of this strategy is the computational cost required to produce the training set, whose size grows quickly with the dimensionality of the parameter space, forcing a trade-off between the range of the sampled cosmologies, the density of the samples in parameter space, and the resolution of the individual simulations.

In this paper, we build a Gaussian-process-based emulator of the HMF, structured in two stages.
A first emulator predicts the overall shape of the HMF given the shape and amplitude of $P_\mathrm{L}(k)$; then, a second emulator corrects the prediction to account for the dependence of the HMF on structure formation history.
Our methodology exploits the evolution mapping framework, which significantly reduces the dimensionality of the input parameter space: consequently, we can train the emulator on a moderate number of simulations and invest more computational resources in their resolution.
Evolution mapping also provides the means to model the non-universal component of the HMF due to structure formation history: we emulate the response of the HMF to changes in formation history and correct the prediction of our baseline emulator with a first-order Taylor expansion.
This emulation strategy is the same as that adopted for the \textsc{Aletheia} emulator of the non-linear matter power spectrum, $P(k)$, in \citet{sanchez_aletheia_2026}, here adapted to the case of the HMF.

This paper is organised as follows.
In Section~\ref{sec:theory}, we summarise the theoretical background on the HMF and its non-universality, and introduce the evolution mapping framework on which our emulation strategy is built.
Section~\ref{sec:methods} describes the design of the two-stage emulator, the suites of N-body simulations used for its training and validation, the estimation of the HMF from the halo catalogues, and the Gaussian process architecture.
In Section~\ref{sec:validation} we validate the emulator against test simulations, demonstrate its applicability to cosmologies beyond $\Lambda\mathrm{CDM}$, and compare its performance with existing emulators.
We summarise our findings and conclude in Section~\ref{sec:conclusions}.

\section{Theory}
\label{sec:theory}

\subsection{HMF background theory}
\label{subsec:HMF}

The differential number density of haloes can be written as
\begin{equation}
    \label{eq:fnu_dndm}
     \dv{n}{\ln M} = \frac{\bar{\rho}}{M} f(\nu) \dv{\ln \nu}{\ln M} \, ,
\end{equation}
where the shape of the HMF is determined by the multiplicity function $f(\nu)$, which in the earliest models of halo collapse \citep[e.g.,][]{press_schechter_hmf_1974, bond_excursion_set_1991} is entirely determined by the peak height
\begin{equation}
    \label{eq:nu_def}
    \nu \equiv \frac{\delta_\mathrm{c}}{\sigma(M, z)} \, .
\end{equation}
All the cosmology- and  time-dependence of the HMF is thus encoded in $\sigma(M, z)$, the RMS fluctuation of the linear density field at redshift $z$, with the field smoothed over a top-hat filter with the halo Lagrangian radius, $R_\mathrm{L} = {(3M / 4 \pi \bar{\rho})}^{1/3}$, while $\delta_\mathrm{c} = 3/5 \, {(3\pi / 2)}^{2/3}$ is the linear overdensity threshold for spherical collapse.

More recent studies of the HMF in simulations have highlighted how a universal form of $f(\nu)$ is not sufficient to describe the HMF to the desired, per cent-level accuracy.
Non-universal features, i.e., additional dependencies of $f(\nu)$ to the peak height, can impact the multiplicity function by over ten per cent \citep[e.g.,][]{jenkins_hmf_2001}.

While many of the existing analytical fitting functions might give a satisfactory effective account of the non-universality, the accuracy degrades significantly at high redshifts or in non-standard cosmologies (\citetalias{fiorilli_hmf_2026}), and they leave the physical drivers of the non-universality unclear.
In the last years, owing to the increase in computational power and consequent improvement of numerical simulations, a number of studies modelled the non-universality of the HMF as a combination of dependencies on two physical effects:
\begin{enumerate}
    \item The shape of the linear matter power spectrum at the scales relevant for halo formation.
    \citet{ondaromallea_hmf_2021}, \citet{castro_euclid_hmf_2023}, and \citetalias{fiorilli_hmf_2026} parametrised this dependency with the effective spectral tilt at halo scales, with marginal differences among the parametrisations;
    \item The growth of structure, which affects the halo density profiles and, in turn, the mass enclosed within a given overdensity threshold, thus changing the HMF. 
    \citetalias{fiorilli_hmf_2026} provided an intuitive physical explanation of such dependency within the evolution mapping framework \citep{sanchez_evo_mapping_2022}, which we detail below.
\end{enumerate}

\subsection{Evolution mapping}
\label{subsec:evmap}
The evolution mapping framework, first formalised in \citet{sanchez_evo_mapping_2022}, is based on splitting the cosmological parameter space into two subsets, shape and evolution parameters, based on the effect different parameters have on the linear matter power spectrum.
\begin{itemize}
    \item \textbf{Shape parameters}, $\vb{\Theta}_{\mathrm{s}}$, controlling the shape of $P_{\mathrm{L}}(k)$. These include the physical baryon and cold dark matter densities, $\omega_{\mathrm{b}}$ and $\omega_{\mathrm{c}}$, and the primordial scalar spectral index, $n_{\mathrm{s}}$;
    \item \textbf{Evolution parameters}, $\vb{\Theta}_{\mathrm{e}}$ that only rescale the amplitude of the linear power spectrum. To this set belong, e.g., the physical dark energy density, $\omega_\mathrm{DE}$, 
    its equation of state parameters, the curvature energy density $\omega_K$ and the primordial power spectrum amplitude $A_\mathrm{s}$. 
\end{itemize}
At the level of the linear power spectrum, the effect of the evolution parameters can thus be summarised into one parameter, expressing the overall clustering amplitude.
Notably, the reduced Hubble parameter $h$, given by the sum of all physical density parameters, $h^2 = \sum_i \omega_i$, mixes shape and evolution and as such, quantities whose value depend on $h$ cannot quantify the overall effect of $\vb{\Theta}_{\mathrm{e}}$.
Hence, we adopt the clustering amplitude $\sigma_{12}$, the RMS fluctuation of the density field smoothed over a spherical top-hat filter of radius 12 Mpc \citep{sanchez_arguments_against_h_2020}.
We can thus write the linear power spectrum as
\begin{equation}
    P_{\mathrm{L}}(k|z,\vb{\Theta}_{\mathrm{s}},\vb{\Theta}_{\mathrm{e}}) = P_{\mathrm{L}}\left(k|\vb{\Theta}_{\mathrm{s}},\sigma_{12}\left(z,\vb{\Theta}_{\mathrm{s}},\vb{\Theta}_{\mathrm{e}}\right)\right).
    \label{eq:pk_evmap_linear}
\end{equation}

The non-linear matter power spectrum is primarily determined by its linear counterpart \citep[e.g.,][]{peacock_dodds_1996}.
Consequently, equation~(\ref{eq:pk_evmap_linear}) holds very well into the non-linear regime \citep[see Figure 2 of][]{sanchez_evo_mapping_2022}, with appreciable deviations only appearing in the highly non-linear regime at late times.
These deviations originate from the cosmology-dependent coupling of the Fourier modes.
In theoretical frameworks developed to describe clustering in the non-linear regime, such as standard perturbation theory \citep[SPT, e.g.][]{bernardeau_pt_2002}, renormalised perturbation theory \citep{crocce_scoccimarro_renormalized_pt_2006}, or the effective field theory of LSS \citep{carrasco_eftoflss_2012, baumann_eftoflss_2012}, the solutions depend on 
\begin{equation}
    x(z) = \frac{\Omega_\mathrm{m}(z)}{f_\mathrm{g}^2(z)} \, ,
    \label{eq:x_def}
\end{equation}
where $\Omega_\mathrm{m}(z)$ is the fractional matter density parameter at redshift $z$ and $f_\mathrm{g}(z) = \dd \ln D(z) / \dd \ln a$ is the linear growth rate\footnote{We added the subscript "g" (for "growth") to the linear growth rate $f_\mathrm{g}(z)$, to avoid any confusion with the multiplicity function $f(\nu)$.}.
The dependence of the coupling kernels on $x$ can often be safely neglected when modelling LSS statistics in the mildly non-linear regime, adopting the constant value $x \equiv 1$, as in an Einstein-de Sitter (EdS) cosmology \citep{Taruya2016, Garny2021}.
However, they need to be accounted for when modelling the matter power spectrum \citep{sanchez_aletheia_2026} or the velocity power spectrum \citep{esposito_evolution_mapping_2024, esposito_velocity_power_spectrum_2026} in the highly non-linear regime.
As the scales at which deviations from the degeneracy of equation~(\ref{eq:pk_evmap_linear}) start manifesting visibly are the scales of haloes ($\sim 1$ Mpc), structure formation history needs to be accounted for when modelling the HMF too (\citetalias{fiorilli_hmf_2026}).

In particular, \citet{sanchez_aletheia_2026} and \citetalias{fiorilli_hmf_2026} showed that this effect can be modelled very effectively with an integrated measure of $x$ over the recent past of history formation, through
\begin{equation}
    \tilde{x}(\tau|\eta) = \int_{-\infty}^\tau d\tau' \, x(\tau) \mathcal{G}(\tau' - \tau | \eta) \, , 
    \label{eq:x_tilde}
\end{equation}
where $\tau = \ln \sigma_{12}$ and $\mathcal{G}$ is a normalised, one-sided Gaussian kernel, whose standard deviation $\eta$ measures the memory of formation history accounted in the integration.
In the fitting function of \citetalias{fiorilli_hmf_2026}, $\eta$ was treated as a free parameter to fit, obtaining different values for each halo mass definition, scaling monotonically with the overdensity threshold $\Delta$ defining the halo border.

The evolution mapping degeneracy, equation~(\ref{eq:pk_evmap_linear}), allows to build accurate emulators of LSS statistics training them on a parameter space of reduced dimensionality.
Given these observables are primarily determined by $P_\mathrm{L}(k)$, one need only train the emulator varying the shape parameters (here, the physical baryon and cold dark matter densities, $\omega_\mathrm{b}$ and $\omega_\mathrm{c}$, and the spectral index $n_\mathrm{s}$), and $\sigma_{12}$. 
Any additional cosmology dependence arising from non-linear evolution can be accounted for with a correction to the emulator prediction as a function of $\tilde{x}$.
We devote the next section to describing in detail the design, training data and architecture of our emulator.

\begin{figure*}
	\centering
	\includegraphics[width=0.95\textwidth]{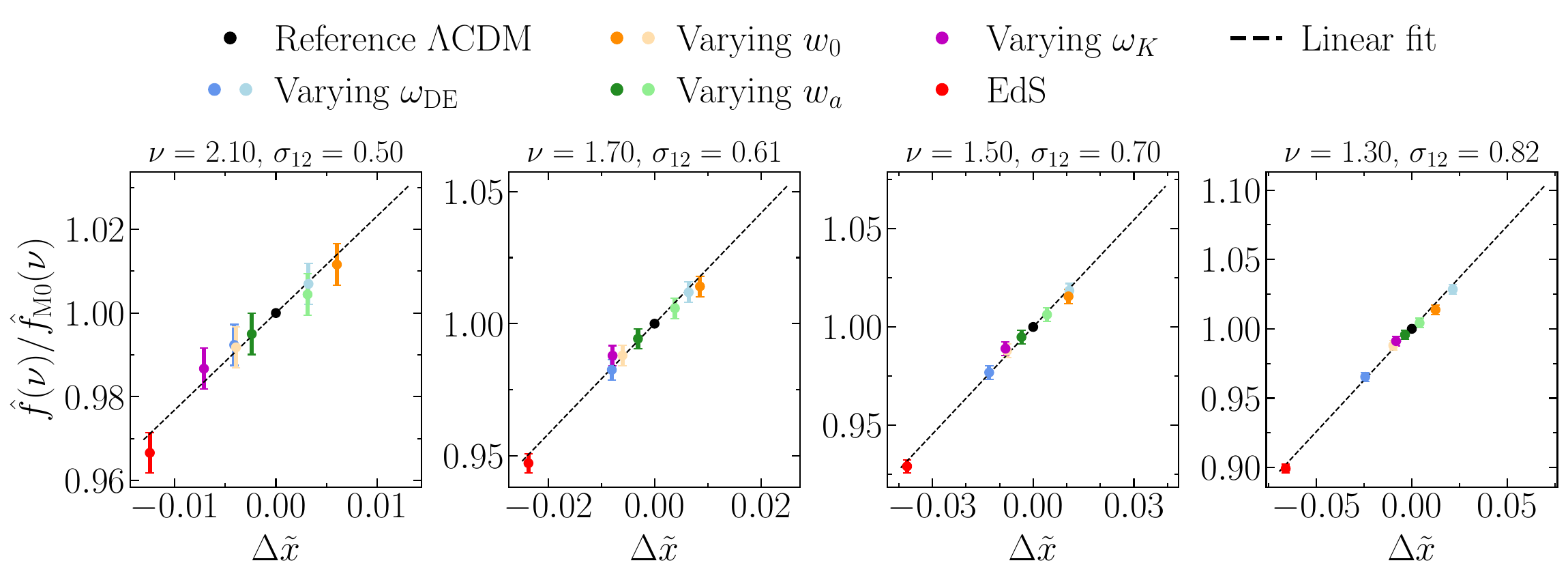}
    \caption{Ratio of the multiplicity function measured in the Aletheia simulations to the measurement in Model 0, our reference $\Lambda \mathrm{CDM}$ cosmology, for an example value of the peak height $\nu$ at each $\sigma_{12}$, as a function of the difference in formation history parameter, $\Delta \tilde{x}$.
    In each panel, the data is fitted with a linear relation constrained to go through $(0,1)$. 
    The remarkable linearity of the response justifies the use of equation~(\ref{eq:R_nu_expansion}) to approximate the response of the HMF to changes in $\tilde{x}$.
    }
    \label{fig:aletheia_response}
\end{figure*}

\section{Methods}
\label{sec:methods}

\subsection{Emulator design}
\label{subsec:design}

Given the dependencies of the HMF discussed in section~\ref{subsec:HMF} and the way we can parametrise the growth history dependence, illustrated in section~\ref{subsec:evmap}, our HMF emulator works in two stages.
First, an emulator predicts the overall shape of $f(\nu)$, while a second emulator accounts for the differences due to structure formation history, parametrised by $\tilde{x}$.
Throughout, we use the notation $\hat{f}(\nu)$ to refer to the estimated multiplicity function, obtained from the binned abundance of the haloes identified in a simulation snapshot, as described in section~\ref{subsec:estimation}.
All the relations below are defined in terms of $f(\nu)$; the emulators are trained on their measured counterparts, in which $f(\nu)$ is replaced by $\hat{f}(\nu)$.

In the first stage, we fix the evolution parameters to a reference set $\vb{\Theta}_{\mathrm{e},0}$. 
Varying only the shape parameters $\vb{\Theta}_\mathrm{s} = \{\omega_\mathrm{b}, \, \omega_\mathrm{c}, \, n_\mathrm{s}\}$ and the clustering amplitude $\sigma_{12}$, we train an emulator to predict the ratio of the multiplicity function to a reference curve $f_\mathrm{ref} (\nu)$:
\begin{equation}
    C(\nu) = \frac{f(\nu|\vb{\Theta}_\mathrm{s}, \vb{\Theta}_{\mathrm{e},0}, \sigma_{12})}{f_\mathrm{ref}(\nu)} \, .
    \label{eq:f_over_fref}
\end{equation}
We take as a reference the fitting function of \citetalias{fiorilli_hmf_2026} evaluated at $z=0$ in a $\Lambda\mathrm{CDM}$ cosmology with parameters fixed at the best-fit values obtained from fits to the CMB temperature and polarisation anisotropy spectra by \citet{planck_2018}. 

In the second stage, we compute the correction needed to account for a set of evolution parameters different from the reference one.
Defining the ratio of the multiplicity functions at a given $\nu$ in two cosmologies with the same $P_\mathrm{L}(k)$, but different formation histories,
\begin{equation}
    R(\nu|\vb{\Theta}_\mathrm{s}, \vb{\Theta}_{\mathrm{e}}, \sigma_{12})) \equiv \frac{f(\nu|\vb{\Theta}_\mathrm{s}, \vb{\Theta}_{\mathrm{e}}, \sigma_{12})}{f(\nu|\vb{\Theta}_\mathrm{s}, \vb{\Theta}_{\mathrm{e},0}, \sigma_{12})} \, ,
    \label{eq:R_nu_def}
\end{equation}
we can expand it in powers of $\Delta\tilde{x} = \tilde{x} - \tilde{x}_0$. 
Truncating at first order, 
\begin{equation}
    R(\nu|\vb{\Theta}_\mathrm{s}, \vb{\Theta}_{\mathrm{e}}, \sigma_{12})) = 1 + \pdv{R}{\tilde{x}}\Big|_{\tilde{x}_0} \Delta \tilde{x} + \mathcal{O}({\Delta\tilde{x}}^2) \, .
    \label{eq:R_nu_expansion}
\end{equation}
If the differences in $\tilde{x}$ are small enough that the truncation of the expansion at first order is sufficient, we can train a second emulator to predict the response $\pdv{R}{\tilde{x}}$ and use equation~(\ref{eq:R_nu_expansion}) to correct the prediction of the first emulator.
In order to test the hypothesis of a linear response, we can look at the ratios of the HMF measured in a set of cosmologies comprising a reference model and variations of its evolution parameters.
These are the Aletheia simulations \citep{esposito_evolution_mapping_2024}. 
They represent nine cosmologies, all sharing the same shape of the linear power spectrum, but varying one evolution parameter each with respect to a reference $\Lambda\mathrm{CDM}$ model, labelled as model 0.
The snapshots are obtained at redshifts such that at any snapshot all models match the same $\sigma_{12}$, so that by construction they all have identical $P_\mathrm{L}(k)$.
Figure~\ref{fig:aletheia_response} shows examples of the ratios of the multiplicity function measured in the Aletheia simulations with respect to that of model 0, for a sample peak height $\nu$ at each snapshot.
The measured ratios are fitted with a linear relation as a function of $\Delta\tilde{x}$, in which we fit the slope and constrain the line to pass through $(0, 1)$, as per equation~(\ref{eq:R_nu_expansion}).
The linear approximation of the response to changes in $\tilde{x}$ thus appears to hold very well as long as $\Delta \tilde{x}$ is of the order of a few $10^{-2}$, as it is in any realistic variation of evolution parameters with respect to a sensible reference cosmology.
Hence, for every training and testing node of the first emulator, we ran simulations varying the evolution parameters, and measured the ratio of $\hat{f}(\nu)$ to the one measured in the original cosmology, fitting it with a linear relation as shown in Fig.~\ref{fig:aletheia_response}.
The second emulator predicts the slope of this linear correction taking as input the same quantities as the first one, $\nu$, $\vb{\Theta}_\mathrm{s}$, and $\sigma_{12}$.
With the output of the two emulators, we construct the full prediction of the multiplicity function as
\begin{equation}
    \begin{split}     
    \mathcal{E}_f(\nu|\vb{\Theta}_\mathrm{s}, \vb{\Theta}_{\mathrm{e}}, \sigma_{12})) &= f_\mathrm{ref}(\nu) \, \mathcal{E}_C(\nu|\vb{\Theta}_\mathrm{s}, \sigma_{12}) \\ &\cdot \Big(1 + \mathcal{E}_{\partial R / \partial \tilde{x}}(\nu|\vb{\Theta}_\mathrm{s}, \sigma_{12}) \, \Delta \tilde{x} \Big) \, .
    \end{split} 
    \label{eq:emu_full}
\end{equation}

\begin{figure}
    \centering
    \includegraphics[width=\linewidth]{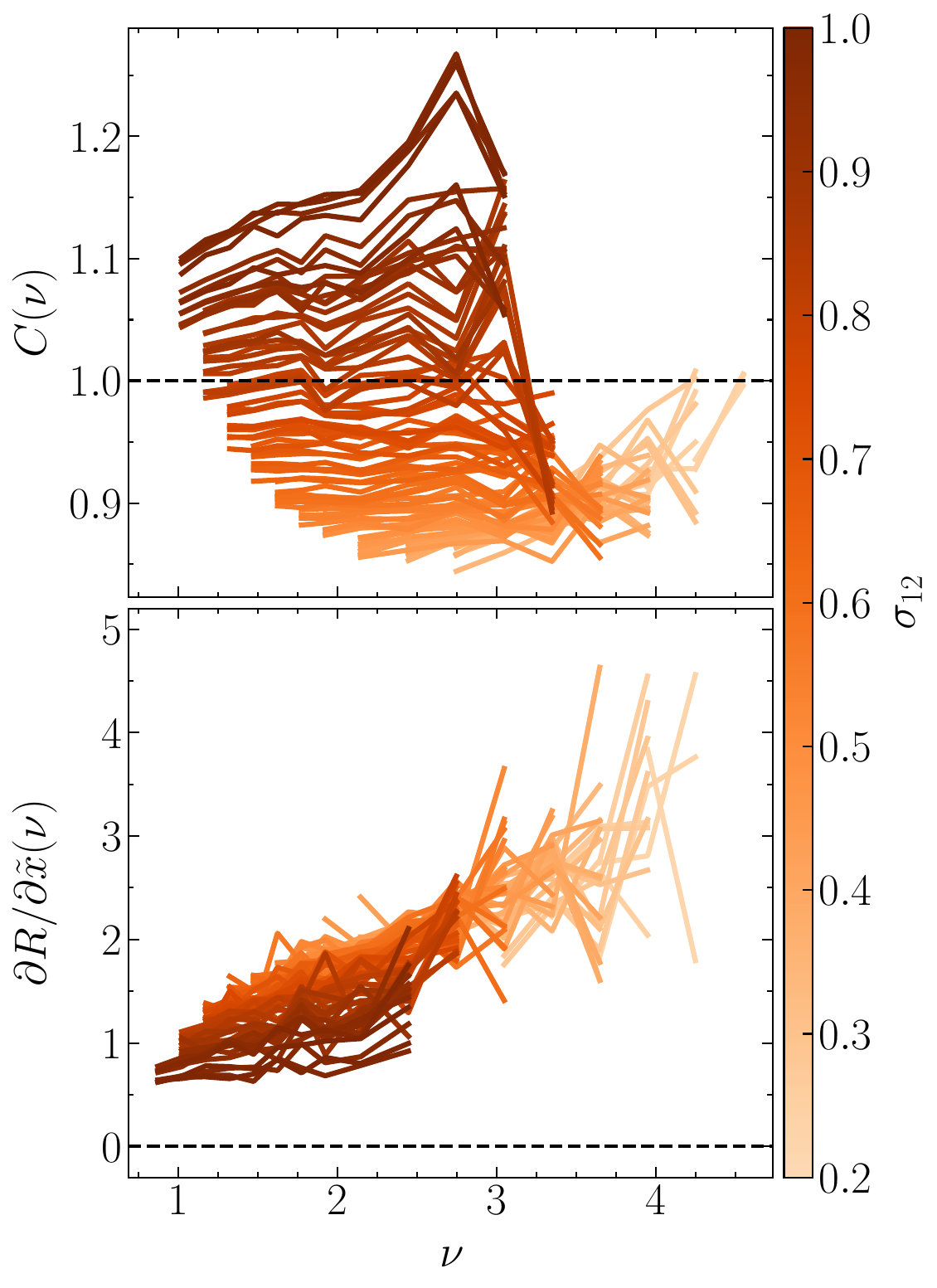}
    \caption{Training data for the two emulators. 
    The upper panel shows the measured multiplicity function in the 100 calibration simulations, divided by the reference curve $f_\mathrm{ref}(\nu)$.
    In the lower panel, the measured response to changes in $\tilde{x}$, obtained by fitting the measurements from the lower-resolution variations of each node.}
    \label{fig:training_data}
\end{figure}

\subsection{Simulations}
\label{subsec:simulations}
We give here a brief description of the cosmological N-body simulations used to train, test and validate the emulators described in this paper.
For a more detailed description, we refer the reader to \citet{sanchez_aletheia_2026}, in which the same simulations were used to build the \textsc{Aletheia} emulator of $P(k)$.

All simulations were run with the publicly available N-body solver \textsc{Gadget4} \citep{springel_gadget4_2021} and 
the initial conditions were set using \textsc{2lptic} \citep{crocce_2lpt_2006,crocce_2lptic_2012} at $z = 99$. 

\begin{itemize}
    \item To train the emulator $\mathcal{E}_C(\nu)$, we use the AletheiaEmu suite, consisting of 100 simulations, all representing flat $\Lambda\mathrm{CDM}$ cosmologies.
    Their parameters cover the nodes of a maximin Latin hypercube \citep{mckay_LH_1979, stein_MLH_1987}.
    Three axes of the hypercube follow the directions of the eigenvectors of the parameter covariance matrix for the shape parameters, $\Theta_{\rm s} = (\omega_\mathrm{b}, \omega_\mathrm{c}, n_\mathrm{s})$, derived from the \citet{planck_2018} CMB temperature anisotropy power spectrum fits.
    These axes are centred on the Planck best-fit values and sampled with a width of $\pm 5$ times the square root of the corresponding eigenvalue.
    The fourth axis is the clustering amplitude $\sigma_{12}$, which is defined independently and sampled from 0.2 to 1.0.
    The resulting parameter-space coverage is shown in Fig. 3 of \citet{sanchez_aletheia_2026}.
    All simulations have a box side length of 1500 Mpc and contain $2048^3$ particles.
    The runs were initialised with paired-and-fixed initial conditions \citep{angulo_pontzen_paired_fixed_2016}, implying each node in the ensemble is composed of two simulations with reversed phases in the initial conditions, and the resulting HMF is the average of those obtained from the two individual boxes.
    We then employ an additional set of 50 AletheiaEmu test simulations, spanning nodes of a different Latin hypercube in the same parameter range, to test the performance of $\mathcal{E}_C(\nu)$.
    The initial conditions for all these simulations were generated with the same random seed; hence they all share the same phases in the Fourier modes of their initial density field (although with different amplitudes, as the linear power spectra are different for each model).

    \item To train the emulator of the response to changes in $\tilde{x}$, $\mathcal{E}_{\partial R / \partial \tilde{x}}(\nu)$, we ran, for each of the AletheiaEmu training and testing nodes, lower-resolution variations changing one evolution parameter, to measure the changes induced from a different $\tilde{x}$ at fixed linear power spectrum, i.e., matching $\vb{\Theta}_\mathrm{s}$ and $\sigma_{12}$. 
    They have the same box size as the original simulations, with $1500^3$ particles.
    We ran one simulation with the same cosmology as the original node to serve as a reference, two with a different dark energy density ($\omega_\mathrm{DE} = 0.925\, \omega_\mathrm{DE}^0$ and $1.125\,  \omega_\mathrm{DE}^0$), and two with non-zero curvature ($\omega_K = -0.04$ and $+0.02$).
    These simulations still feature initial conditions generated from the power spectrum fixed to its mean value, without sampling a realisation of cosmic variance, but they are not "paired", i.e., each node consists of only one realisation.
    All simulations used to train our emulators have the same box size, and similar mass resolutions.
    We verify in Appendix~\ref{app:resolution_test} that, when tested against simulations with different size and resolution, the accuracy of the prediction is not impacted.
    
    \item To validate the performance of our emulator and demonstrate its applicability beyond $\Lambda\mathrm{CDM}$ cosmologies, we use an additional suite of simulations, named AletheiaDE.
    This suite varies both shape and evolution parameters, including the dark energy equation of state parameters $w_0$ and $w_a$.
    It comprises a reference $\Lambda\mathrm{CDM}$ model and 10 different $w_0w_a\mathrm{CDM}$ cosmologies, one of which corresponds to the best-fit parameters from the DESI DR1 BAO data combined with CMB anisotropy spectra and Union3 supernovae luminosity distances \citep{DESI_Y1_BAO_cosmology_2025}.
    The specific parameter values for these models are detailed in Table 1 of \citet{sanchez_aletheia_2026}.
    These runs match the resolution of our primary training suite, use paired-and-fixed initial conditions, and output at six redshifts: $z = 3, \, 2, \, 1, \, 0.6, \, 0.3$, and 0.
    While all AletheiaDE realisations share the same initial phases, these are chosen to be different from the phases used in the AletheiaEmu calibration set.
   
\end{itemize}

\subsection{Measuring the HMF}
\label{subsec:estimation}

In all our simulations, we identify haloes with the \textsc{Rockstar} halo finder \citep{behroozi_rockstar_2013}.
We employ the default setting of excluding from the identified haloes the particles that, despite lying within the halo radius, are not gravitationally bound to it.
We remove in post-processing all subhaloes, also identified by \textsc{Rockstar}, whose centre lies within the radius of their parent halo.
We define our haloes with an overdensity threshold of 200 times the comoving background matter density, $\Delta = \rho_\mathrm{halo} / \bar{\rho} = 200$.
The same emulator design, architecture and training procedure may be applied, unchanged, to other overdensity thresholds.

\begin{figure}
    \centering
    \includegraphics[width=\linewidth]{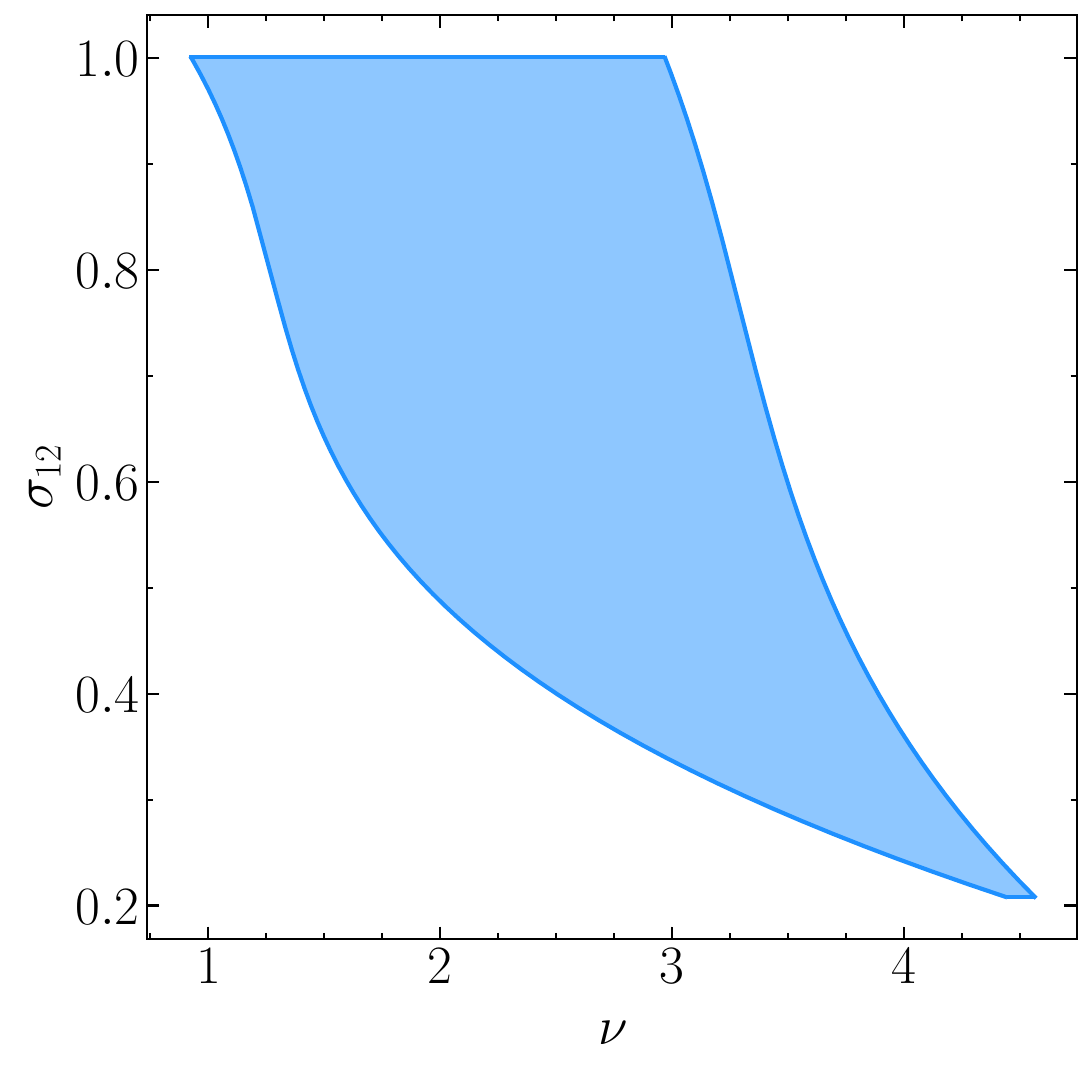}
    \caption{Region of the $(\nu, \sigma_{12})$ plane spanned by the training set of the shape emulator $\mathcal{E}_C$. 
    At each value of the clustering amplitude $\sigma_{12}$, the accessible range of peak height $\nu$ is set by the mass resolution and number count cuts described in Section~\ref{subsec:estimation}: the low-$\nu$ edge corresponds to the minimum mass threshold, while the high-$\nu$ edge follows from the minimum number of haloes required per bin. 
    Lower clustering amplitudes shift the accessible window towards higher $\nu$, as a given mass maps to a larger peak height.}
    \label{fig:param_range}
\end{figure}

We bin the halo catalogues with a linear binning scheme in $\nu$, with $\Delta \nu = 0.15$ for $\nu \in [0.8, \, 2)$, and $\Delta \nu = 0.3$ for $\nu \in [2,\, 5]$.
To ensure convergence of the HMF, we take a minimum mass threshold of 1000 particles per halo to construct the catalogues used to train $\mathcal{E}_{C}(\nu)$, and consider as our lowest-$\nu$ bin the one whose low-mass edge is the closest to such mass.
To avoid training the emulators on measurements with too big a scatter due to Poisson noise, we take as highest-$\nu$ bin the last one containing more than a threshold number of haloes, which we set to 100 for training $\mathcal{E}_C(\nu)$.
The area covered by the projection of the training set in the $(\nu, \sigma_{12})$ plane is shown in Fig.~\ref{fig:param_range}. 
The accessible peak height range at each clustering amplitude is a direct consequence of the cuts we just introduced: the lower-$\nu$ edge corresponds to the minimum mass threshold, while the upper-$\nu$ edge is set by the requirement of a minimum number of haloes per bin. 
Since a given mass maps to a larger peak height at lower $\sigma_{12}$, the accessible window shifts towards higher $\nu$ as the clustering amplitude decreases. 
When constructing the training set for $\mathcal{E}_{\partial R / \partial \tilde{x}}(\nu)$, we find that at very high $\nu$ the scatter of the measured responses is much higher than that of $C(\nu)$.
To ensure having a meaningful training set for the emulator to learn from, we increase the minimum number of haloes per bin to 1000 in the catalogues used to measure $\pdv{R}{\tilde{x}}$.
As a result, the response emulator $\mathcal{E}_{\partial R / \partial \tilde{x}}(\nu)$ is trained on a slightly narrower area of the $(\nu, \sigma_{12})$ plane than $\mathcal{E}_{C}(\nu)$.
In the area only covered by the training data of $\mathcal{E}_{C}(\nu)$, we find that letting the response emulator work in extrapolation does not significantly impact the overall accuracy of the prediction.
We thus take the training set bounds of $\mathcal{E}_{C}(\nu)$, as limits of applicability of the full emulator.

We estimate the binned multiplicity function, $\hat{f}(\nu)$, summing the halo masses directly in the estimator, rather than assigning a representative bin mass, to avoid the bin miscentering effect highlighted in \citet{Li_Smith_2025}. The estimator takes the form (\citetalias{fiorilli_hmf_2026}):
\begin{equation}
    \hat{f}(\nu) = \frac{1}{\bar{\rho} V_\mathrm{box}} \frac{1}{\Delta \ln \nu} \sum_{\nu_i \in [\nu_1, \nu_2)} M_i \, ,
    \label{eq:fnu_binned_estimator}
\end{equation}
for a bin of low- and high-mass edges $\nu_1$ and $\nu_2$, respectively, where $\Delta \ln \nu = \ln (\nu_2 / \nu_1)$ is the logarithmic bin width and $V_\mathrm{box} = {L_\mathrm{box}}^3$ is the volume of the simulated box.

We compute the uncertainties on $\hat{f}(\nu)$  following \citet{smith_marian_counts_covariance_2011}, accounting for the Poisson noise in the halo counts and a sample variance term due to the finite size of the simulation.

When computing the integrated formation history parameter $\tilde{x}$, we employ the value of the memory parameter $\eta$ obtained in \citetalias{fiorilli_hmf_2026} fitting the HMF at $\Delta = 200$, $\eta = 0.288$

\subsection{Architecture}
\label{subsec:architecture}
Each of our two emulators is a Gaussian process (GP) regressor.
A GP is a collection of random variables, any finite subset of which follows a joint multivariate Gaussian distribution \citep{rasmussen_williams_gp_2006}.
The joint distribution is fully specified by a mean function, which can be taken to be zero after standardising the training data, and a covariance function, or kernel.
The kernel $K(\vb{x}, \vb{x'})$ is a real-valued, symmetric and positive semi-definite function defined over the same domain as the functions that constitute the GP: in our case, the domain is the input parameter space $(\vb{\Theta}_\mathrm{s}, \sigma_{12}, \nu)$.
The kernel encodes the characteristic scales of variation of the target function with respect to the input parameters, and the smoothness of the target function.

A GP provides a prediction of its target function as follows: we construct a subset of the GP with $N$ training observations of the target function $\vb{y} = \{y_i(X=\vb{x}_i)\}$, $i \in \{1, \dots, N\}$, together with $M$ unobserved realisations $\vb{y'} = \{y'_j(X'=\vb{x'}_j\}$, $j \in \{1, \dots, M\}$, which we want to predict given their input parameters $\vb{x'}_j$.
Their joint distribution is assumed to be a multivariate Gaussian, with zero mean and covariance
\begin{equation}
    \label{eq:gp_covariance}
    K = \begin{bmatrix}
        \hat{K}_{XX} & K_{XX'} \\
        K_{X'X}      & K_{X'X'}
    \end{bmatrix} ,
\end{equation}
where each block is the covariance of the respective input parameters, and the one relative to the training data includes the uncertainty on the observed realisations, $\hat{K}_{XX} = K(X,X) + \mathrm{diag}(\sigma_1^2, \ldots, \sigma_N^2)$, where $\sigma_i^2$ is the variance of the $i$-th observation $y_i$, estimated as described in Section~\ref{subsec:estimation}.
We can condition over the first $N$ variables to get a conditional posterior distribution $\mathcal{P}(\vb{y'}|X', \vb{y}(X))$.
The conditional posterior is still a multivariate Gaussian \citep{rasmussen_williams_gp_2006}, with mean and covariance given by
\begin{align}
    \vb{\mu'} &= K_{X'X} \, {(\hat{K}_{XX})}^{-1} \, \vb{y} \, , \label{eq:gp_posterior_mean}\\
    K' &= {({K^{-1}}_{X'X'})}^{-1} \, . \label{eq:gp_posterior_covariance}
\end{align}
The posterior mean of equation~(\ref{eq:gp_posterior_mean}) is the emulator prediction, while the posterior covariance of equation~(\ref{eq:gp_posterior_covariance}) provides a built-in estimate of the emulation uncertainty.
The ability of GPs to interpolate accurately from a relatively modest number of training points makes them particularly well suited to emulation problems in which each training sample is expensive to produce, as is the case for N-body simulations.

For both emulators, we employ a Matérn kernel with a smoothness parameter of $\nu_\mathrm{M} = 3/2$\footnote{The smoothness parameter of the Matérn kernel, only taking half-integer values, is generally indicated with $\nu$. We added the subscript "M" to avoid confusion with the peak height.}.
This kernel takes the form
\begin{equation}
    K_{\nu_\mathrm{M} = 3/2} = \sigma_f^2 \Big(1 + \sqrt{3} r\Big) \exp \Big(-\sqrt{3} r\Big) \, .
    \label{eq:matern_kernel}
\end{equation}
The distance $r$ between two points in the input parameter space, $\vb{x}$ and $\vb{x'}$, is given by the anisotropic Mahalanobis metric
\begin{equation}
    r(\vb{x}, \vb{x'}) = \sqrt{\sum_i {\Bigg(\frac{x_i - x'_i}{\ell_i}\Bigg)}^2} \, .
    \label{eq:mahalanobis_metric}
\end{equation}
The hyperparameters to be optimised during the training process are the signal variance ${\sigma_f}^2$ and the characteristic length scales of each input dimension, $\ell_i$.
Allowing for a separate length scale $\ell_i$ along each input dimension, a choice known as automatic relevance determination, lets the training process learn how rapidly the target function varies with each input: dimensions with little influence on the output are assigned large length scales and are effectively marginalised over.

The hyperparameters $(\sigma_f, \{\ell_i\})$ are fixed by maximising the log marginal likelihood of the training data,
\begin{equation}
    \ln p(\vb{y}|X) = -\frac{1}{2} \vb{y}^\mathrm{T} \, \hat{K}_{XX}^{-1} \, \vb{y} - \frac{1}{2} \ln \abs{\hat{K}_{XX}} - \frac{N}{2} \ln 2\pi \, ,
    \label{eq:gp_marginal_likelihood}
\end{equation}
where the dependence on the hyperparameters enters through the training covariance matrix $\hat{K}_{XX}$ of equation~(\ref{eq:gp_covariance}).
The first term rewards the goodness of fit to the training data, while the second penalises model complexity, so that their combination provides a degree of protection against overfitting without the need for a separate validation set \citep{rasmussen_williams_gp_2006}.
As the marginal likelihood surface can be multimodal, the optimisation is restarted from several random initial positions in hyperparameter space, and the solution with the highest marginal likelihood is retained.
To train the emulators, we used the Python package \texttt{scikit-learn} \citep{pedregosa_scikitlearn_2011}.

\section{Validation}
\label{sec:validation}

\subsection{Validation of the individual emulators}
\label{subsec:test_sims}

\begin{figure}
    \centering
    \includegraphics[width=\linewidth]{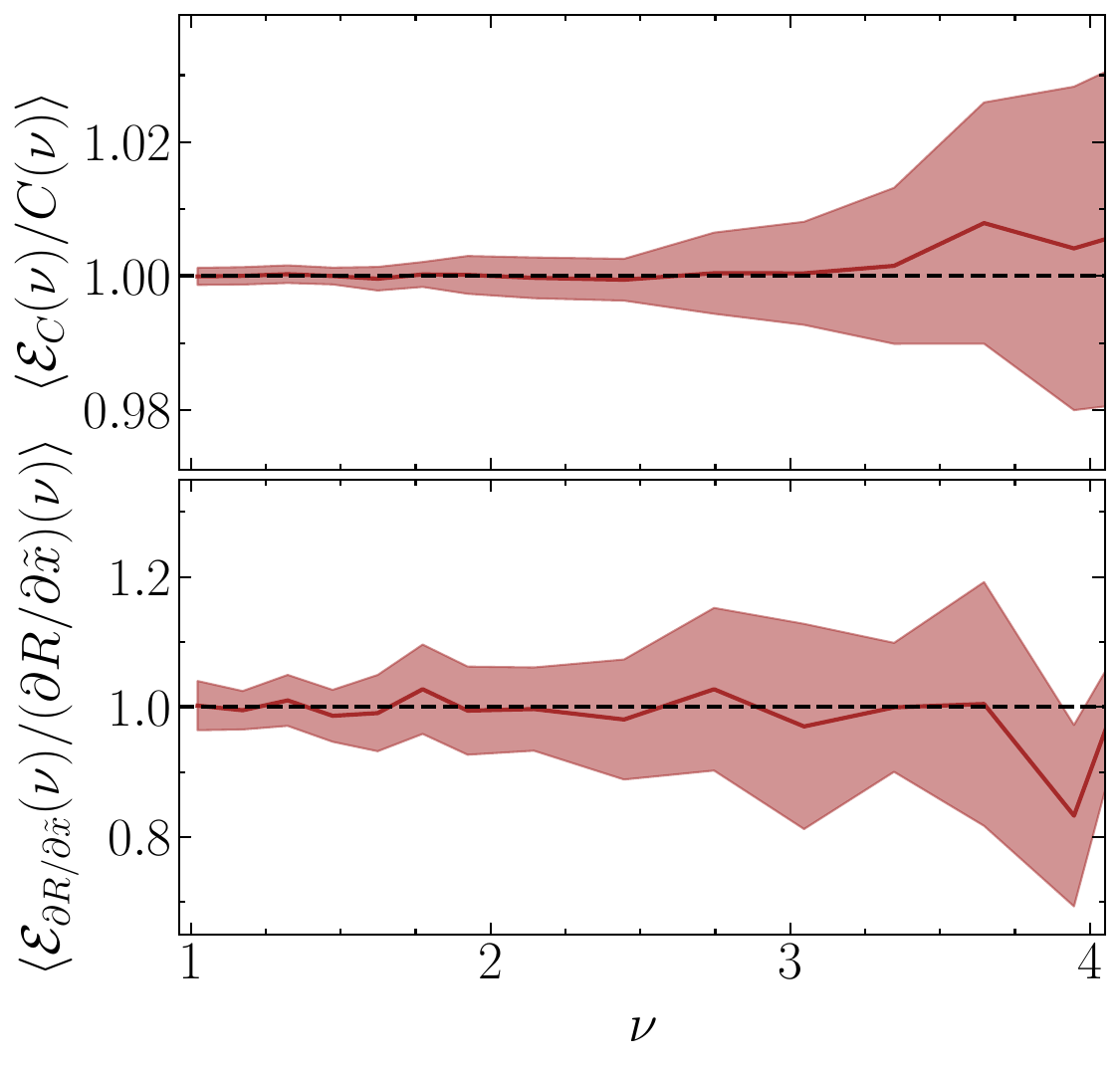}
    \caption{Performance of the two emulators on the AletheiaEmu testing simulation set.
    The upper panel shows the accuracy of the emulator of $C(\nu)$ and the lower panel that of the emulator of the response to evolution differences, $\partial R / \partial \tilde{x}$.
    In both panels, the solid line and shaded area show the mean and the standard deviation of the ratio of the emulator prediction to the respective measured quantity in the test simulations.}
    \label{fig:performace_separate}
\end{figure}

We first validate the two emulators separately, on simulations not used in their training. 
For the emulator of the shape of $f(\nu)$, $\mathcal{E}_C(\nu)$, we use the 50 $\Lambda\mathrm{CDM}$ test simulations spanning an independent Latin hypercube in $(\vb{\Theta}_\mathrm{s}, \sigma_{12})$, as described in Section~\ref{subsec:simulations}; for the response emulator $\mathcal{E}_{\partial R / \partial \tilde{x}}(\nu)$, we use the derivative simulations run at the corresponding test nodes. 
Figure~\ref{fig:performace_separate} shows, for each emulator, the averaged ratio of the predicted functions to the values measured in the test simulations, with the solid line and shaded band denoting the mean and the standard deviation across the test set.

The shape emulator (upper panel) is remarkably accurate: its mean prediction is unbiased at well below the per-cent level across the entire peak-height range, and the scatter remains below one per cent up to $\nu \simeq 3$, exceeding it only in the rare-halo tail, where the measurements themselves become noisier owing to the small number of haloes per bin.
The response emulator (lower panel) is also accurate, but it exhibits a larger scatter, increasing towards high $\nu$.
This degradation is expected, as the response is measured from lower-resolution simulations, by taking ratios of their mass functions and fitting the derivative of the ratios with respect to $\tilde{x}$.
Each of these steps propagates measurement noise, which is amplified in the sparsely populated high-$\nu$ bins, where the emulator has learnt only from a handful of responses. 
This loss of precision at high $\nu$, however, has a negligible impact on the final HMF prediction. 
The response enters equation~(\ref{eq:emu_full}) multiplied by $\Delta\tilde{x}$, which is itself of order few $10^{-2}$ for any realistic departure from the reference evolution parameters $\vb{\Theta}_{\mathrm{e},0}$. 
In particular though, such large peak heights are relevant for the HMF at high redshift (see Fig.~\ref{fig:param_range}), and at those epochs $\Delta\tilde{x}$ is smaller still, generally of order $10^{-3}$, because all cosmologies approach an Einstein--de Sitter behaviour during matter domination, so that they all converge to $\tilde{x} = 1$. 
The high-$\nu$ region where the response emulator is least precise is therefore the region where its contribution to the total prediction is most strongly suppressed.

\subsection{Accuracy in $\Lambda$CDM and comparison with public emulators}
\label{subsec:lcdm_accuracy}

\begin{figure*}
    \centering
    \includegraphics[width=\textwidth]{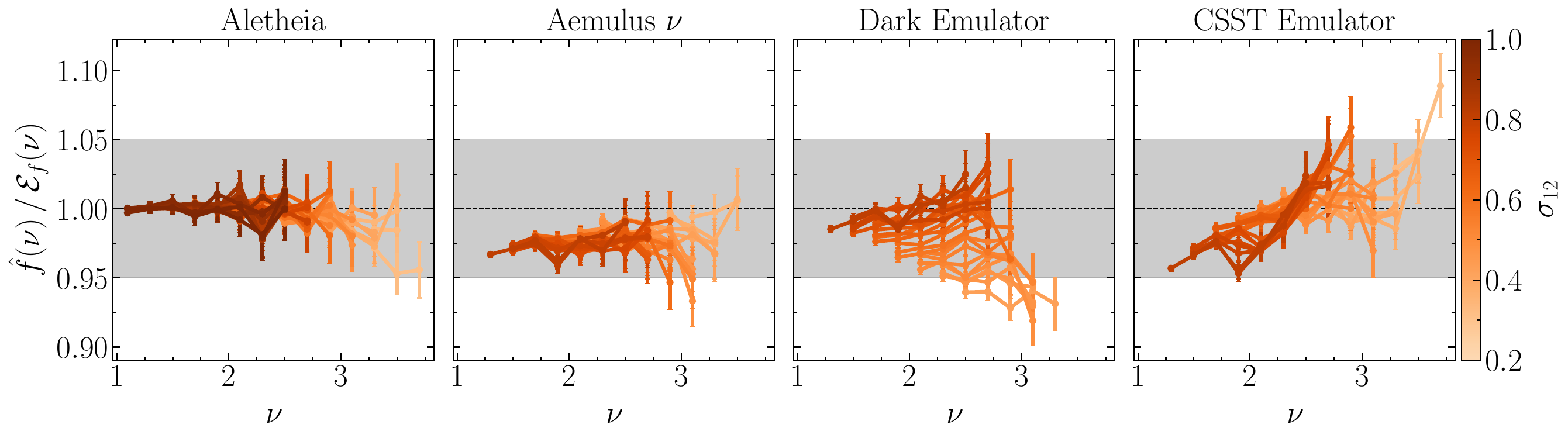}
    \caption{Multiplicity function measured in the AletheiaEmu test simulations divided by the prediction of, from left to right: our emulator, the \textsc{Aemulus~$\nu$} emulator \citep{shen_aemulusnu_hmf_emulator_2025}, the \textsc{Dark Emulator} \citep{nishimichi_darkquest_emulator_2019}, and the \textsc{CSST Emulator} \citep{chen_yu_CSST_emulator_2025}.
    Each line corresponds to one test simulation and the colour encodes its clustering amplitude $\sigma_{12}$.
    The shaded band marks the $\pm5$ per cent range. 
    Each external emulator is evaluated only on the test set nodes whose redshifts fall within its training domain.}
    \label{fig:comparison_lcdm}
\end{figure*}

A growing number of HMF emulators have been released in recent years (Section~\ref{sec:intro}).
A fully like-for-like comparison is difficult because they adopt different halo mass definitions, cover different regions of cosmological parameter space, and are validated over different redshift ranges.
On a flat $\Lambda\mathrm{CDM}$ footing, however, we can use the 50-node $\Lambda\mathrm{CDM}$ AletheiaEmu test set we introduced in Section~\ref{subsec:simulations} to compare our emulator of the HMF shape, $\mathcal{E}_C (\nu)$, against three public emulators: \textsc{Aemulus~$\nu$} \citep{shen_aemulusnu_hmf_emulator_2025}, \textsc{Dark Emulator} \citep{nishimichi_darkquest_emulator_2019} and the \textsc{CSST Emulator} \citep{chen_yu_CSST_emulator_2025}.
All three are built from \textsc{Rockstar} catalogues with the $\Delta = 200$ mass definition.
We evaluate all four emulators on the test set, restricting each external emulator to the redshifts that fall within its own training domain: $0<z<2$ for \textsc{Aemulus~$\nu$}, $0<z<1.48$ for the \textsc{Dark Emulator}, and $0 < z < 3$ for the \textsc{CSST Emulator}.
Figure~\ref{fig:comparison_lcdm} shows the ratio of the multiplicity function measured in the test simulations to the emulator predictions, colour-coded by the clustering amplitude $\sigma_{12}$.
Our emulator is unbiased over nearly all the peak-height range, with deviations approaching the few-per-cent level only in the rare-halo tail at low $\sigma_{12}$, where the measurements themselves are noisiest, as seen in Fig.~\ref{fig:performace_separate}.
\textsc{Aemulus~$\nu$} systematically overpredicts the results of our test simulations by  2--3 per cent.
The \textsc{Dark Emulator} provides accurate predictions at late times, with residuals increasing with redshift (i.e., with decreasing $\sigma_{12}$) up to 5--6 per cent.
The \textsc{CSST Emulator} displays a $\nu$-dependent tilt in the ratios of $\hat{f}(\nu)$ to its predictions, overestimating the abundance by up to $4$ per cent at low peak heights and underestimating it by a comparable amount in the rare-halo tail.
Note that, as the test simulations share the same initial density field phases as our training set, this comparison reduced the scatter in the predictions of $\mathcal{E}_C (\nu)$. 
Evaluating the emulator on simulations with independent initial phases would introduce sample variance and increase the scatter of these ratios. 
However, this would not induce a systematic bias; the lack of a systematic offset confirms the accuracy of the emulator's mean predictions.

\subsection{Validation of the full emulator beyond $\Lambda$CDM}
\label{subsec:extensions}

As a result of the evolution mapping degeneracy, equation~(\ref{eq:pk_evmap_linear}), using the clustering amplitude expressed in terms of $\sigma_{12}$ and correcting the emulated HMF for the differences in structure formation history in terms of $\tilde{x}$ is sufficient to emulate $f(\nu)$ in cosmologies with virtually any combination of evolution parameters, provided $\sigma_{12}$ and the shape parameters $\vb{\Theta}_\mathrm{s}$ remain within the training range.
This includes cosmologies with non-zero curvature, quintessence and dynamical dark energy models.

\begin{figure*}
    \centering
    \includegraphics[width=\textwidth]{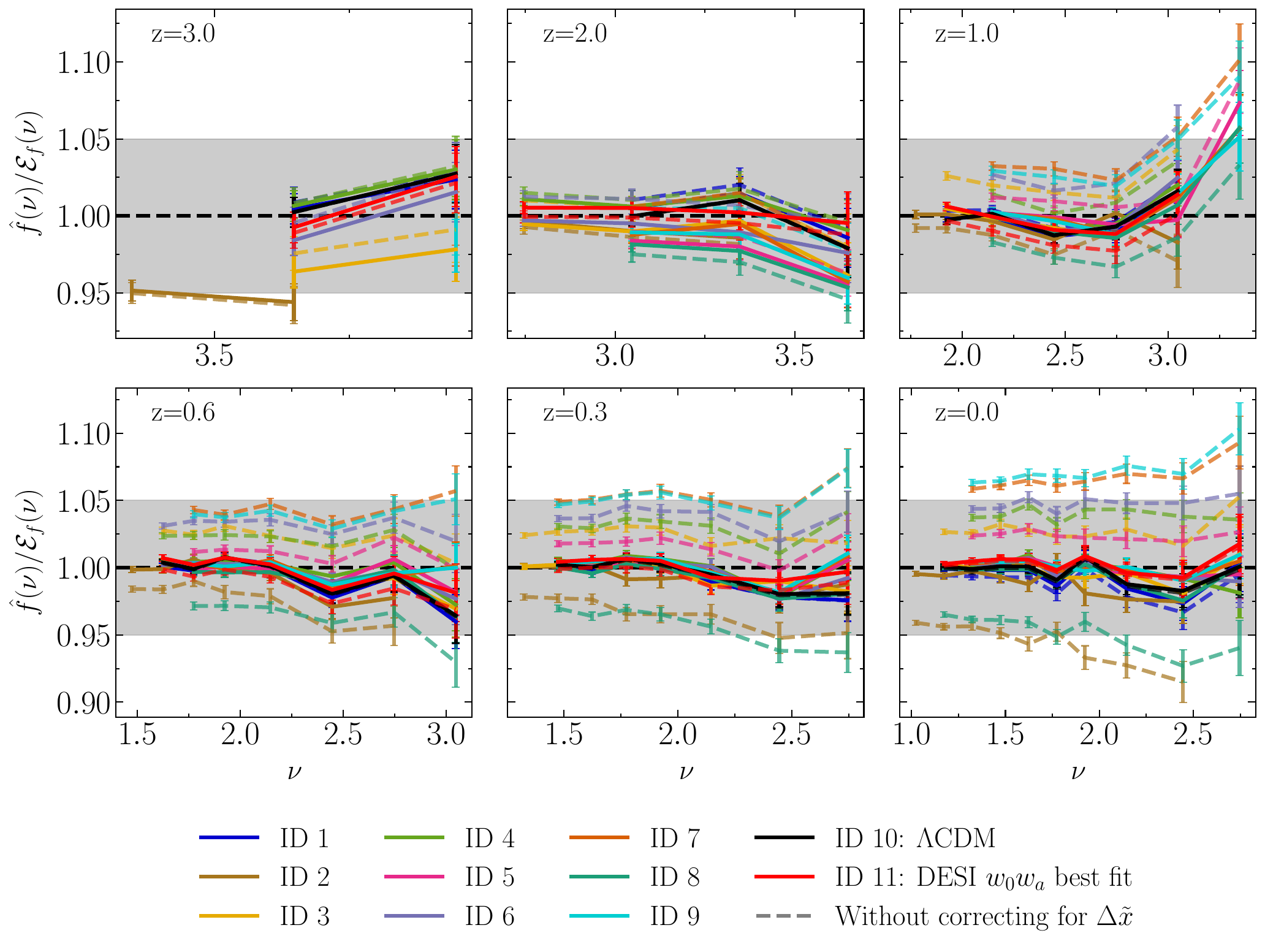}
    \caption{Multiplicity function measured in the AletheiaDE simulations divided by the prediction of the full emulator (solid lines) and of the emulator without the structure-formation-history correction (dashed lines).
    Each panel corresponds to one of the six output redshifts, and each colour to one of the simulated cosmologies: the reference $\Lambda\mathrm{CDM}$ model (ID 10), the DESI Y1 $w_0 w_a$ best-fit cosmology (ID 11), and nine further $w_0 w_a \mathrm{CDM}$ models (ID 1--9). 
    The values of the cosmological parameters of each model can be found in Table 1 of \citet{sanchez_aletheia_2026}.
    The shaded band marks the $\pm5$ per cent region.}
    \label{fig:withoutx}
\end{figure*}

To demonstrate this point, we validate the full emulator on the AletheiaDE simulations.
These simulations feature 11 models in the $w_0w_a\mathrm{CDM}$ parameter space, varying both shape and evolution parameters, including the best-fit cosmology of DESI Y1 BAO+CMB+SN \citep{DESI_Y1_BAO_cosmology_2025}, and one reference $\Lambda\mathrm{CDM}$ model. 
Figure~\ref{fig:withoutx} shows the ratio of the multiplicity function measured in these simulations to the prediction of our emulator (solid lines).
The residuals are nearly always below five per cent, demonstrating the applicability of the emulator to the full evolution-parameter space, larger-dimensional than that spanned by the training simulations, at fixed coverage of the shape parameters and $\sigma_{12}$.

To show the role of the response emulator $\mathcal{E}_{\partial R / \partial \tilde{x}}$, the same figure compares the residuals of the full emulator with those obtained when the correction is omitted, setting $\Delta \tilde{x} = 0$ in equation~(\ref{eq:emu_full}) (dashed lines). 
At low redshifts, where the differences in $\tilde{x}$ between cosmologies are sizeable, the residuals broaden significantly when the correction is not applied. 
At high redshift, the two predictions nearly coincide, as all cosmologies approach an Einstein--de Sitter behaviour and $\Delta\tilde{x} \to 0$. 
This demonstrates that our first-order correction in $\Delta\tilde{x}$ captures the non-universality imprinted by structure formation history effectively.

The AletheiaDE simulations were run with initial phases independent of those of the AletheiaEmu suite, providing a test free from the phase correlation discussed in Section~\ref{subsec:lcdm_accuracy}. 
The ratios shown as solid lines in Fig.~\ref{fig:withoutx} remain at the same level of scatter as those shown in the leftmost panel Fig.~\ref{fig:comparison_lcdm}. 
This demonstrates that the emulator does not systematically overfit to the training phases, and that the added scatter from sample variance across different realisations is small compared to the intrinsic interpolation error of $\mathcal{E}_C(\nu)$.

\subsection{Comparison with models valid beyond $\Lambda$CDM}
\label{subsec:comparisons}

\begin{figure*}
    \centering
    \includegraphics[width=\textwidth]{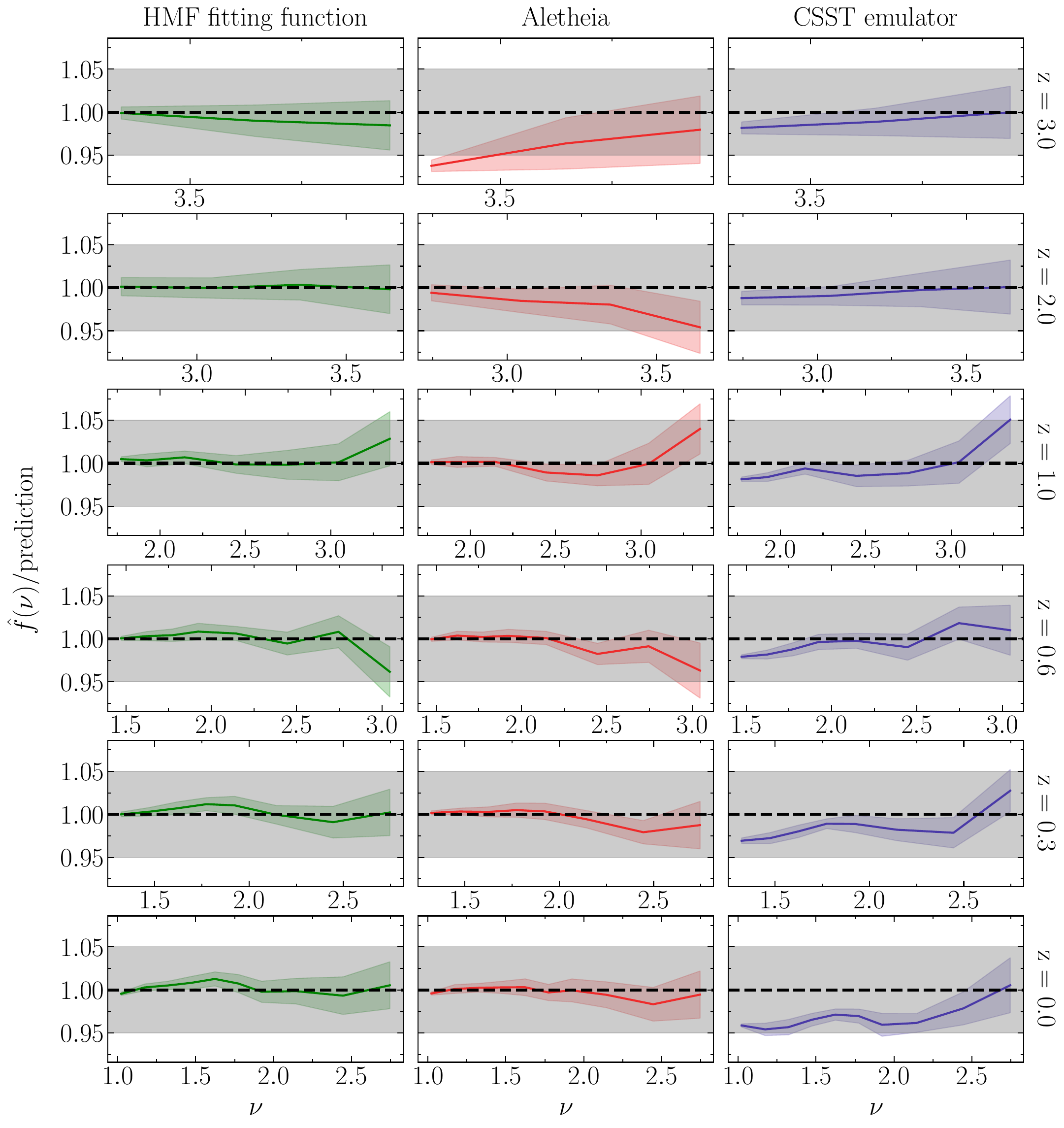}
    \caption{Comparison of our emulator (central column) to the fitting function of \citetalias{fiorilli_hmf_2026} (left column) and the \textsc{CSST Emulator} \citep[][right column]{chen_yu_CSST_emulator_2025}.
    The solid lines and coloured shaded areas show the mean and the standard deviation of the ratio of the multiplicity function measured in the AletheiaDE simulations to the respective predictions.
    Each row corresponds to one of the six output redshifts. 
    The grey shaded bands mark the $\pm5$ per cent region in each panel.}
    \label{fig:comparison_emutest_Aletheia_CSST_fit}
\end{figure*}

In order to benchmark the performance of the emulator beyond $\Lambda$CDM cosmologies against previously existing tools, we compare it against the \textsc{CSST Emulator} \citep{chen_yu_CSST_emulator_2025} and the fitting function of \citetalias{fiorilli_hmf_2026}.
Among the three emulators considered for the comparison shown in Fig.~\ref{fig:comparison_lcdm}, the \textsc{CSST Emulator} is the only one trained on cosmologies with dynamical dark energy, thus applicable to predict the HMF from the AletheiaDE simulations.
Figure~\ref{fig:comparison_emutest_Aletheia_CSST_fit} shows the averaged ratios of multiplicity functions measured in the AletheiaDE simulations divided by the prediction of these three tools: the \citetalias{fiorilli_hmf_2026} fitting function in the left column; our emulator in the central one; and the \textsc{CSST Emulator} in the right column.
Each row corresponds to one snapshot of the simulations, and the shaded bands indicate the standard deviation of the ratios across models.
The fitting function of \citetalias{fiorilli_hmf_2026} provides an analytic description of the same non-universal HMF but requires an explicit functional form, calibrated on simulations spanning different combinations of evolution parameters, including a pair of $w_0w_a\mathrm{CDM}$ models.
The accuracy of the fitting function on data from these simulations is very similar to that of the emulator.
The \textsc{CSST Emulator} performs equally well, and even slightly better on some models at high redshift, but it shows a slight bias at low redshift and low masses, which becomes a nearly constant offset of a few per cent in the last snapshot, at $z=0$.
Note that two models of the AletheiaDE, including the DESI Y1 $w_0w_a$ best-fitting cosmology, lie outside the range of validity of the \textsc{CSST Emulator}, and are thus not included in the right column panels.
As our emulator extends beyond $\Lambda\mathrm{CDM}$ purely by means of its correction accounting for the response to different structure formation history, its range of validity beyond-$\Lambda$CDM evolution parameters is virtually unlimited.

\section{Conclusions}
\label{sec:conclusions}

We have presented a Gaussian process emulator of the halo mass function, built within the evolution mapping framework \citep{sanchez_evo_mapping_2022} and following the two-stage strategy introduced by \citet{sanchez_aletheia_2026} to build the \textsc{Aletheia} emulator of the non-linear matter power spectrum. 
The emulator predicts the multiplicity function $f(\nu)$ for haloes defined at an overdensity threshold $\Delta = 200$. 
A first stage of the emulation, $\mathcal{E}_C$, predicts the shape of $f(\nu)$ as a function of the shape parameters $\vb{\Theta}_\mathrm{s} = \{\omega_\mathrm{b}, \omega_\mathrm{c}, n_\mathrm{s}\}$ and the clustering amplitude $\sigma_{12}$.
Then, a second emulator, $\mathcal{E}_{\partial R / \partial \tilde{x}}$, corrects the prediction to account for the dependence on structure formation history through a first-order expansion in $\tilde{x}$. 
Because evolution mapping collapses the dependence on all evolution parameters into $\sigma_{12}$ and $\tilde{x}$, the emulator can be trained on a low-dimensional parameter space sampled by only 100 high-resolution $\Lambda\mathrm{CDM}$ simulations, and their low-resolution variations to measure $\partial R / \partial \tilde{x}$, while remaining applicable to cosmologies with any combination of evolution parameters, e.g., curvature, quintessence and dynamical dark energy.
The resulting emulator is publicly available in the latest release of the {\sc Aletheia} emulator suite\footnote{The \textsc{Aletheia} Python package is available on PyPI at \url{https://pypi.org/project/AletheiaCosmo/}. Code and documentation are available at \url{https://gitlab.mpcdf.mpg.de/arielsan/aletheia}.}.

We validated the two stages separately on independent test simulations: the shape emulator is unbiased at well below the per-cent level, while the response emulator is noisier at high peak height, in a regime where its contribution to the full prediction is strongly suppressed.
On our flat $\Lambda\mathrm{CDM}$ test set, where the evolution parameters are fixed to the reference set and the response correction is therefore inactive, the shape emulator alone reproduces the measured HMF without bias, whereas \textsc{Aemulus~$\nu$} \citep{shen_aemulusnu_hmf_emulator_2025} and \textsc{Dark Emulator} \citep{nishimichi_darkquest_emulator_2019} carry a $2$--$4$ per cent offset, while the \textsc{CSST Emulator} \citep{chen_yu_CSST_emulator_2025} exhibits a slight $\nu$-dependent tilt in the ratios of measured to predicted multiplicity functions. 

We then tested the performance of the full emulator on beyond-$\Lambda$CDM cosmologies, employing the AletheiaDE $w_0 w_a \mathrm{CDM}$ simulations.
The emulator predicts the HMF to an accuracy of a few per cent across six redshifts and a wide range of evolution parameters, matching the accuracy of the fitting function of \citetalias{fiorilli_hmf_2026}. 
The \textsc{CSST Emulator} \citep{chen_yu_CSST_emulator_2025} attains similar accuracy at high $\nu$, while it gives slightly biased predictions for low peak heights at low redshifts, remaining valid over a smaller portion of parameter space than our emulator.

The results of this work highlight the broad applicability of the evolution mapping framework.
The strategy of splitting the cosmological dependence into shape parameters, $\sigma_{12}$, and a  first-order correction in $\tilde{x}$ previously used to build our $P(k)$ emulator can be successfully extended to other statistics of the highly non-linear density field.
By compressing the impact of all evolution parameters into a single amplitude parameter, we retain the advantage of training the emulator on a lower-dimensional parameter space, while accurately capturing residual differences in structure formation history through the $\tilde{x}$ correction. 
Moving forward, this result opens up the possibility of expanding the Aletheia suite to encompass a comprehensive set of cosmological statistics based on the same unified theoretical principle.

\section*{Acknowledgements}
We would like to thank Carlos Correa, Sofia Contarini, Matteo Esposito, Alejandro Pérez Fernández, Johan Comparat, and Barbara Sartoris for their help and useful comments.
The simulations used in this work, Aletheia, AletheiaEmu, AletheiaDE and AletheiaMass were carried out and post-processed on the HPC systems Cobra, Raven and Viper of the Max Planck Computing and Data Facility (MPCDF, \url{https://www.mpcdf.mpg.de}) in Garching, Germany.
This project has received funding from the European Union’s HORIZON-MSCA-2021-SE-01 Research and Innovation programme under the Marie Skłodowska-Curie grant agreement number 101086388 - Project acronym: LACEGAL.

\section*{Data Availability}

The \textsc{Aletheia} emulator of the HMF described in this work is publicly available and registered on the Python Package Index (PyPI). 
The simulation data on which this paper is based can be shared on reasonable request to the authors.


\bibliographystyle{mnras}
\bibliography{bibliography}



\appendix

\section{Robustness to simulations of varying resolution}
\label{app:resolution_test}

All simulations used to train the emulator have the same box size (Section~\ref{subsec:simulations}). 
While the number of simulated particles is different between the set used to train $\mathcal{E}_C(\nu)$ and those used for $\mathcal{E}_{\partial R / \partial \tilde{x}}(\nu)$, the mass resolutions of the two sets only differ by a factor of $\sim 2.5$.
We must verify whether this feature of the training set of simulations affects the accuracy of the predictions, when tested against simulations of different resolutions.
In fact, the \textsc{Aletheia} power spectrum emulator of \citet{sanchez_aletheia_2026}, on which our methodology is based, did require an explicit resolution correction.

\begin{figure*}
    \centering
    \includegraphics[width=0.9\textwidth]{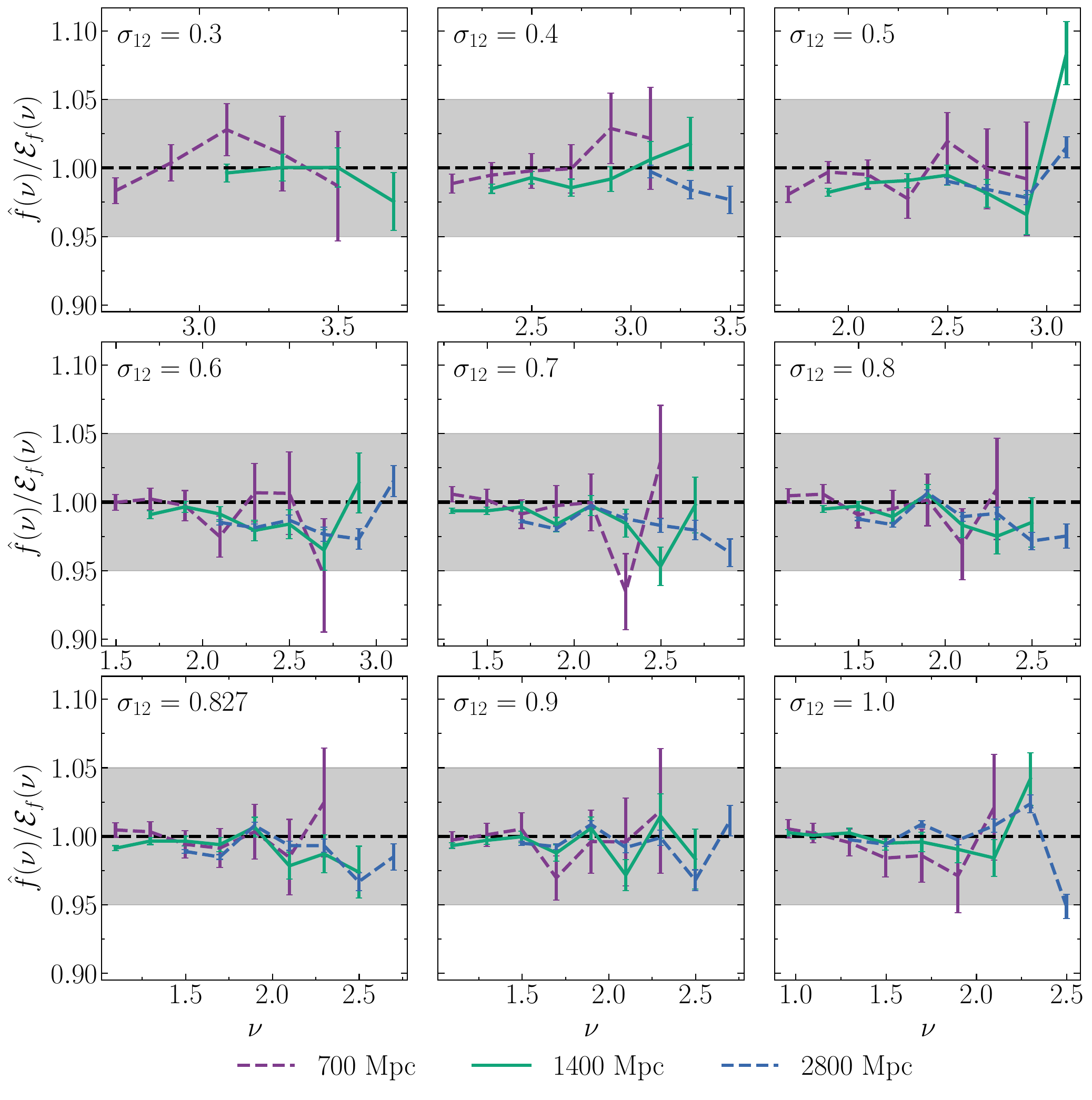}
    \caption{Multiplicity function measured in different boxes of the AletheiaMass simulations (\citetalias{fiorilli_hmf_2026}) divided by the emulator prediction. The simulation of box side length 1400 Mpc (green solid lines) has very similar resolution to the ones used to train the emulator. The data from the other two boxes, of side length 700 and 2800 Mpc (purple and blue dashed lines, respectively) show how the emulator accuracy is insensitive to the resolution of the simulations. Each panel corresponds to one value of $\sigma_{12}$, and the shaded band marks the $\pm5$ per cent region.}
    \label{fig:resolution_test}
\end{figure*}

To test the resolution dependence of our HMF emulator, we use the AletheiaMass simulations (\citetalias{fiorilli_hmf_2026}).
Specifically, we use the boxes of side length 700, 1400 and 2800 Mpc which, at a fixed particle number of $2048^3$, span a factor of 64 in particle mass.
The 1400 Mpc box has a resolution close to that of our training simulations, while the 700 and 2800 Mpc boxes are, respectively, eight times finer and eight times coarser in mass. Figure~\ref{fig:resolution_test} shows the multiplicity function measured in each box, at nine values of $\sigma_{12}$, divided by the emulator prediction.

The three boxes agree with the emulator within the estimated uncertainties and, importantly, do not show any systematic trend with resolution.
The emulator therefore requires no resolution correction. 
This robustness follows from the convergence of the HMF of haloes resolved with at least 1000 particles (Section~\ref{subsec:estimation}), and is consistent with the convergence tests underlying other HMF emulators, which report per-cent-level convergence for haloes sampled with several hundred to about a thousand particles \citep{mcclintock_aemulus_hmf_emulator_2019, bocquet_miratitan_hmf_emulator_2020}.

\bsp	
\label{lastpage}
\end{document}